\documentstyle[aps,epsf,rotate,multicol]{revtex}

\begin{document}

\setlength{\textheight}{24.5cm}
\draft

\newcommand{\figdir}{.}

\title{Universality classes for rice-pile models}

\author{ 
Lu\'{\i}s~A.~Nunes Amaral$^{(1,2),}$\thanks{E-mail: amaral@cmt0.mit.edu} 
and Kent B{\ae}kgaard Lauritsen$^{(3),}$\thanks{E-mail: kent@nbi.dk} 
}

\address{
$^{(1)}$ Theory II, Institut f\"ur Festk\"orperforschung, Forschungszentrum 
	J\"ulich, D-52425 J\"ulich, Germany \\
$^{(2)}$ Physics Dept., Condensed Matter Theory, Massachusetts Institute
	of Technology, Cambridge, MA 02139\\
$^{(3)}$ Niels Bohr Institute, Center for Chaos and Turbulence Studies,
        Blegdamsvej 17, 2100 Copenhagen \O, Denmark
}

\date{submitted: September 27, 1996; printed: \today}

\maketitle

\begin{abstract}

  We investigate sandpile models where the updating of unstable
  columns is done according to a stochastic rule. We examine the
  effect of introducing nonlocal relaxation mechanisms.  We find that
  the models self-organize into critical states that belong to three
  different universality classes.  The models with local relaxation
  rules belong to a known universality class that is characterized
  by an avalanche exponent $\tau \approx 1.55$, whereas the
  models with nonlocal relaxation rules belong to new universality
  classes characterized by exponents $\tau \approx 1.35$ and $\tau
  \approx 1.63$.  We discuss the values of the exponents in terms of
  scaling relations and a mapping of the sandpile models to interface
  models.

\end{abstract}


\begin{multicols}{2}

The concept of self-organized criticality (SOC), proposed in a seminal
paper by Bak, Tang and Wiesenfeld \cite{BTW}, has made considerable
impact on a number of fields in the natural and social sciences.  The
paradigm of SOC is an idealized sandpile where grains added to a pile
dissipate their potential energy through avalanches with no
characteristic scale \cite{BTW,KNWZ,CL,FF,Toner,KSG,PO,Frette,HK}.
Early experimental studies of real sandpiles lead to clear
disagreement with the numerical models: bounded distributions of
avalanche sizes were observed instead of the expect power law
dependence \cite{JLN,H,RVK,RVR,Feder}.  On the other hand, recent
rice-pile experiments found power-law distributions
of avalanche sizes \cite{Rice} and tracer transit times
\cite{christensen-etal:1996}.  These results sparked a renewed
interest in the study of sandpile models
\cite{christensen-etal:1996,amaral-lauritsen:1996a,amaral-lauritsen:1996b,
Hernan,MSorensen,luebeck-usadel:1993,Gerald,nakanishi-sneppen:1996}.

A class of sandpile models with stochastic toppling rules,
which is denoted ``rice-pile'' models, was found to display SOC
in one dimension with a power-law distribution of avalanche sizes
characterized by the critical exponent $\tau \approx 1.55$
\cite{christensen-etal:1996,amaral-lauritsen:1996a,
amaral-lauritsen:1996b,Hernan,MSorensen}. 
Recently,
Ref.~\cite{paczuski-boettcher:1996} proposed a mapping of the model in
\cite{christensen-etal:1996} to the motion of a linear interface
through a disordered medium \cite{heiko}.  For this universality
class, to which we will refer to as the local linear interface (LLI)
universality class, the mapping allows the determination of all the
exponents characterizing the dynamics of the pile
\cite{paczuski-boettcher:1996,us},
and shows that $\tau$ is clearly different from the mean-field value
$\tau=3/2$ (see, e.g., \cite{ZLS}). 
Several other SOC
models have been conjectured to be in the LLI class
\cite{nakanishi-sneppen:1996,paczuski-boettcher:1996,manna,zaitsev}.
Thus, the question arises of what mechanisms
lead to the emergence of the LLI universality class for one-dimensional
stochastic sandpile models with a preferred direction.
Previously, sandpile models in higher dimensions have been classified 
according to their degree of 'directedness'
\cite{KNWZ,Biham}.

In this paper
we undertake an investigation of the mechanisms
responsible for the emergence of a given universality class for
one-dimensional 
sandpile models with stochastic toppling rules, and we discuss the
reasons for the robustness of the LLI universality class.  To this
end, we study a class of models with stochastic toppling rules.
We use as our basic model the one we proposed in
Ref.~\cite{amaral-lauritsen:1996a} and investigate the robustness of
the critical behavior upon modification of the toppling process.
Specifically, we study stochastic variants of the models proposed in
\cite{KNWZ}.

We find that the local models belong to the LLI universality class and
have $\tau \approx 1.55$, while the nonlocal models belong to {\it
  new\/} universality classes with $\tau \approx 1.35$ and $\tau
\approx 1.63$.  In order to understand the new exponent values, we
discuss scaling relations fulfilled by the exponents and the mapping
to interface models.

First, we define the class of one-dimensional models: The system
consists of a plate of length $L$, with a wall at $i = 0$ and an open
boundary at $i = L+1$.  The profile of the pile evolves through two
mechanisms: deposition and relaxation.  Deposition is always done at
$i = 1$, and one grain at a time.  During relaxation we look at all
{\it active\/} columns of the pile: A column $i$ of the pile is
considered active if, in the previous time step, it (i) received a
grain from column $i-1$, for the local models, or from $i-j$,
$j=1,2,...N$, for the nonlocal models, (ii) toppled a grain to column
$i+1$, or (iii) column $i+1$ toppled one grain to its right neighbor.
If a column $i$ is active {\it and\/} the local slope \mbox{$\delta
h_i \equiv h(i) - h(i+1) > S_1$}, then with probability $p(\delta
h_i)$ several grains will be toppled from column $i$.  Here, we study
the case
\begin{equation}
        p(\delta h_i) = \min\left\{ 1, g (\delta h_i - S_1) \right\},
        \label{e-p}
\end{equation}
where $g \le 1$ is a parameter \cite{other_functions}.  The number of
grains $N$ to be toppled is determined by either the {\it limited\/}
($l$) or {\it unlimited\/} ($u$) rule \cite{KNWZ}:
\begin{equation}
        N = \left\{ 
                \begin{array}{ll}
                  N_0  ,              &  ~~~~~(l)  \\
                  \delta h_i - S_1  , &  ~~~~~(u)  
                \end{array}
            \right.
                                        \label{eq:N}
\end{equation}
The toppled grains are then redistributed either according to the
{\it local\/} ($L$) or {\it nonlocal\/} ($N$) rule \cite{KNWZ}:
\begin{equation}
           \begin{array}{ll}
              h(i+1)=h(i+1) + N  ,                   &  ~~~~~(L)  \\
              h(i+j)=h(i+j) + 1,  \qquad j=1,..., N, &  ~~~~~(N)
           \end{array}
                                        \label{eq:h(i+1)}
\end{equation}
Grains toppled to columns $i>L$ leave the system.  We designate the
models by the two letters that describe the rules applied: the local
limited and unlimited models are referred to as $Ll$ and $Lu$, while
the nonlocal limited and unlimited models are referred to as $Nl$ and
$Nu$.  The parameter $p(\delta h_i)$ describes friction between the
grains and accounts for the fact that a large range of ``stable''
slopes was observed in the rice-pile experiment.  The parameter $g$
accounts for the effect of gravity on the packing configurations.  For a
discussion of the physical interpretation of the models
see also Refs.\ \cite{amaral-lauritsen:1996a,amaral-lauritsen:1996b}.

\begin{figure}[ht]
\narrowtext
\centerline{
\epsfysize=0.85\columnwidth{\rotate[r]{\epsfbox{\figdir/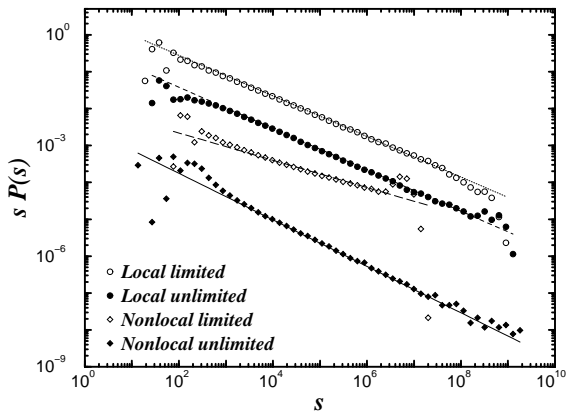}}}
}
\vspace*{0.5cm}
\caption{
  Log-log plot of the probability density of the dissipated
  potential energy $s$ during an avalanche.  For greater clarity, the
  data for the $Lu$, $Nl$, and $Nu$ was divided by factors of $10^1$,
  $10^2$ and $10^3$, respectively.  $P(s)$ was multiplied by $s$ to
  make visually more clear the change in the exponent $\tau$.  For each
  model, we performed runs with the order of $10^7$ grains deposited.
  The results shown are for $L = 1600$, $g = 1/8$, $S_1 = 6$, and $N_0
  = 2$ (for the limited models).  Other values of $S_1$ (ranging from
  1 to 6), of $1/g$ (ranging from 4 to 8), or of $N_0$ (ranging from 2
  to 4) were also investigated without any observable change in the
  estimate of the exponents.  The slopes of the straight lines
  correspond to the exponent values from Table \protect\ref{tab_exp}.
  For the $Nl$ model the bump before the cutoff leads to finite size
  corrections, and we carried out simulations on systems up to size
  $L=20000$ in order to obtain more accurate estimates for the
  exponents.
}
\label{fig1} 
\end{figure}

We study the models in the slowly driven limit, i.e., we take the rate
of deposition to be slow enough that any avalanche, which might be
started by a deposited particle, will have ended before a new particle
is deposited.  The simulations of the models show that each system
quickly enters a steady state which is characterized by varying
avalanche sizes and a complex structure in time
\cite{amaral-lauritsen:1996a}.  The size $s$ of an avalanche can be
defined in a number of ways.  First, we follow the definition of Ref.\ 
\cite{Rice}, and calculate the size of an avalanche as the total
potential energy dissipated in between deposited particles.  In
Fig.~\ref{fig1} we show the avalanche distributions for the various
models. We find that the probability density function can be well
described by the power law form
\begin{equation}
        P(s, L) = s^{-\tau} f(s/L^\nu),
                                        \label{eq:P}
\end{equation}
where $\tau$ and $\nu$ are critical universal exponents.
(Alternatively, we have that $P(s,L) = L^{-\beta} \tilde{f}(s/L^\nu)$,
where $\beta=\nu \tau$.)
We used plots of consecutive slopes for different system sizes to obtain
the estimates for $\tau$ (see table~\ref{tab_exp}).
Then, these values were used to collapse the
data according to Eq.~(\ref{eq:P}) and extract $\nu$.

The results from our numerical simulations show that the $Ll$ and $Lu$
models belong to the LLI universality class with $\tau \approx 1.55$
and $\nu \approx 2.25$, in agreement with the results of
Refs.~\cite{christensen-etal:1996,amaral-lauritsen:1996a}.  On the
other hand, for the {\it nonlocal\/} models we obtain values of the
exponents that signal the existence of {\it new\/} universality
classes. The nonlocal limited ($Nl$) model is characterized by the
exponents $\tau = 1.35 \pm 0.05$ and $\nu=1.55 \pm0.05$. The combined
change of the number of toppled particles together with a nonlocal
relaxation leads to yet a new universality class: For the nonlocal
unlimited ($Lu$) model, we obtain the exponent values $\tau = 1.63 \pm
0.02$ and $\nu=2.75 \pm 0.05$.  We note that simple power laws, as in
Eq.~(\ref{eq:P}), in all cases provide us with nice data collapses.
This result should be contrasted with the investigation of similar
rules for the (nonstochastic) models in Ref.~\cite{KNWZ} where the
results could not be described by a simple power law but instead
required a multifractal scaling form.

\begin{figure}[hb]
\narrowtext
\centerline{
\epsfysize=0.85\columnwidth{\rotate[r]{\epsfbox{\figdir/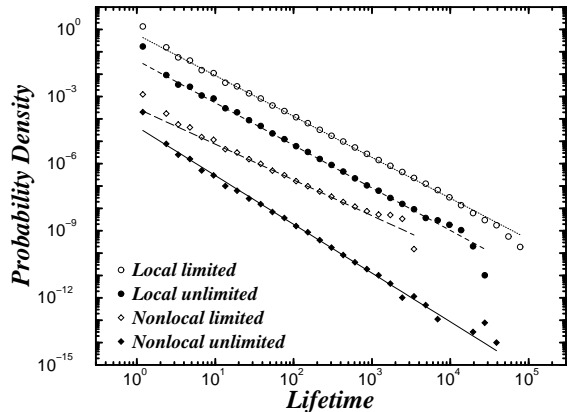}}}
}
\vspace*{0.5cm}
\caption{
  Log-log plot of the probability density for the lifetime of
  avalanches.  For greater clarity, the data for the $Lu$, $Nl$, and
  $Nu$ was divided by factors of $10$, $10^3$ and $10^4$,
  respectively.  The same values were used for the parameters as in
  Fig.~\protect\ref{fig1}.  The slopes of the straight lines
  correspond to the exponent values from Table \protect\ref{tab_exp}.
}
\label{fig2}
\end{figure}

To further test our conclusion regarding the universality classes for
the rice-pile models, we study different definitions of avalanche
size: When using the total number of topplings, we find the same
values for $\tau$ and $\nu$ as quoted above, which is due to the fact
that a toppling event on average dissipates a fixed amount of
potential energy (for a bounded distribution of slopes).  For the
lifetime $T$, we find that the probability density function is well
described by the scaling form
\begin{equation}
        D(T, L) = T^{-y} g(T/L^\sigma),
                                        \label{eq:D}
\end{equation}
with the exponent values listed in table \ref{tab_exp} (see also
Fig.~\ref{fig2}).  These results reassure us that the $Ll$ and $Lu$
models indeed belong to the LLI universality class, whereas the $Nl$
and $Nu$ models belong to new universality classes.

Next, we discuss scaling relations that are obeyed by the critical
exponents.  It is well established that the average avalanche size
scales as
\begin{equation}
        \left< s \right> \sim L^q ,
                                        \label{eq:<s>}
\end{equation}
with the value of the exponent $q$ depending on how the pile is driven.
Here we have $q=1$ for all models (see,
e.g., \cite{KNWZ} and \cite{nakanishi-sneppen:1996} for other $q$
values).  Combining Eqs.~(\ref{eq:P}) and (\ref{eq:<s>}), we obtain
the exponent relation \cite{KNWZ}
\begin{equation}
        \nu (2 - \tau) = q ,
                                        \label{eq:relation1}
\end{equation}
which is fulfilled by all models.  We have used $\left< s \right>
\sim L^{\nu (2 - \tau)}$ (in combination with $\left< s^2
\right>$) to obtain alternative estimates of the critical
exponents, and we obtain values in complete accordance with those
listed in table~\ref{tab_exp}.
Similar relations to Eqs.\ 
(\ref{eq:<s>})--(\ref{eq:relation1}) can be derived for the lifetimes
regarding the averages $\left< T \right>$, $\left< T^2 \right>$, and
then used to extract the critical exponents. Notice, however, that for
the $Nu$ model, $\left< T \right>$ is a constant since we have $y>2$.
{}From the scaling of the probability densities (\ref{eq:P}) and
(\ref{eq:D}), we obtain
\begin{equation}
        \nu (\tau - 1) = \sigma (y - 1) ,
                                        \label{eq:relation2}
\end{equation}
in agreement with our results ($s \sim T^\omega$, with
$\omega=\nu/\sigma$).  The above relations imply that there are only two
independent exponents.

As noted above, it was shown in
Ref.~\cite{paczuski-boettcher:1996} that the rice-pile model in
\cite{christensen-etal:1996} can be mapped to a linear interface model
described by the continuum equation
\begin{equation}
  \frac{\partial H}{\partial t} = D \frac{\partial^2 H}{\partial x^2}
  + \eta(x,H).
                                        \label{eq:int}
\end{equation}
Here, $H$, which is obtained as $h(x,t)=H(x-1,t)-H(x,t)$, counts the
number of topplings of a given column and $\eta(x,H)$ is a quenched
``noise'' which is related to the stochastic toppling probability.
The rice-pile dynamics imposes a driving of the interface at $x=0$.
The mapping predicts that $\nu = 1+ \chi$, where $\chi$ is the
so-called roughness exponent characterizing, e.g., the scaling of the
width of the interface with system size $L$ (see, e.g.,
\cite{laszlo}).  Numerically one has $\chi \approx 1.25$ \cite{heiko}
in excellent agreement with the value $\nu \approx 2.25$.  In
addition, the cutoff exponent for the avalanche lifetime is $\sigma =
z$, where $z$ is the so-called dynamic exponent which describes the
propagation of correlations.

The results presented here show that the $Ll$ and $Lu$ models belong
to the LLI universality class. This result can be easily understood
from the mapping to the linear interface model: The toppling of
several particles does not change the fact that the growth of the
interface is still local.  Since the surface tension term is
the relevant term (in the renormalization group sense) it follows that
the coarse-grained behavior of the models is governed by the interface
equation (\ref{eq:int}) as for the models in
Refs.~\cite{christensen-etal:1996,amaral-lauritsen:1996a}.

\begin{figure}
\narrowtext
\centerline{
\epsfysize=0.85\columnwidth{\rotate[r]{\epsfbox{\figdir/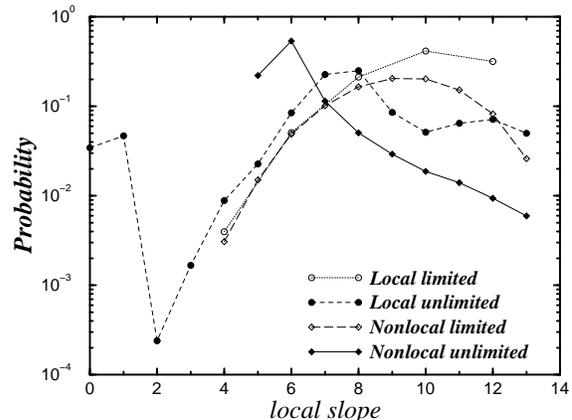}}}
}
\vspace*{0.5cm}
\caption{
  Linear-log plot of the probability of the local slopes
  in the steady state.  The
  parameters of the runs are the same as in Fig.~\protect\ref{fig1},
  except that the system size is $L = 400$.  It is visually apparent,
  that while the limited models lead to truncated Gaussian
  distributions, the unlimited models lead to more complicated forms.
  In particular, the $Nu$ model leads to a nearly exponential decay of
  the probability of finding slopes larger than $S_1$ (which is
  related to the increase in the value of $\tau$ for this model).
}
\label{fig3}
\end{figure}

On the other hand, for the $Nl$ and $Nu$ models, the nonlocal toppling
rules generate nonlocal growth which affect the interface motion.  It
has previously been shown that nonlocal interactions will in general
lead to the emergence of new universality classes
\cite{bray:1990,lauritsen:1995} which is confirmed by our observations
for the $Nl$ and $Nu$ models.  The critical exponents for the
interface equations corresponding to the $Nl$ and $Nu$ models are not
known.  We can, nevertheless, qualitatively understand the change in
the values of the exponent $\tau$ as follows: In general, we would
expect that a nonlocal toppling rule would lead to a decrease of the
value of $\tau$ because more columns in the pile are perturbed at
every time step, thus creating larger avalanches.  This is indeed what
happens for the
$Nl$ model. However, for the $Nu$ model we observe an
increase in the value of $\tau$.  To understand this result, it is
useful to look at the average slope of the pile and at the
distribution of local slopes for the different models (see
Fig.~\ref{fig3}).  For the limited models, the average slope remains
practically the same upon the change of the number of particles
toppled: 9.3 for the $Ll$ model and 9.1 for the $Nl$ model.  This
means that the change from one (local) to many (nonlocal) columns
being ``actived'' works as we described above, i.e., it leads to an
lower value of $\tau$.  On the other hand, for the unlimited models, a
large change in the average slope is observed when we change the
number of particles toppled: 7.8 for the $Lu$ model and 6.2 for the
$Nu$ model.  This implies that, on average, only a few particles are
toppled from unstable columns for the $Nu$ model, but for the
$Lu$ model more particles are toppled and this leads to an higher
likelihood of big avalanches.  As a result, we conclude that the $Nu$
model should have a higher value of $\tau$ than the $Lu$ model, as is
indeed observed.

In summary, we study a class of rice-pile models which in a simple way
model some of the physical features of the experiments in Refs.\
\cite{Rice,christensen-etal:1996}.  We find that for local relaxation
rules the models belong to the LLI universality class which is
characterized by the exponents $\tau \approx 1.55$ and $y \approx
1.87$ (models $Ll$ and $Lu$).  On the other hand, for models with
nonlocal relaxation rules we obtain more complex dynamics resulting in
new universality classes: For the $Nl$ model we obtain $\tau \approx
1.35$ and $y \approx 1.60$, while for the $Nu$ model we obtain $\tau
\approx 1.63$ and $y \approx 2.20$. 
Our results show that nonlocal rules can increase or decrease the value
of $\tau$, thus opening for the possibility that the rice-pile
experiment in Ref.\ \cite{Rice} can be explained by some extension of
nonlocal sandpile models by incorporating in detail the rice-grain
dynamics.

We acknowledge discussions with A. Corral, M. H. Jen\-sen, J. Krug, M.
Mar\-ko\-sova, and K. Sneppen.  K.B.L. thanks the Danish Natural
Science Research Council for financial support.

\begin{table}[b]
\narrowtext
\caption{ Critical exponents for the four models studied.
  The definition of the critical exponents and classification of the
  models are given in the text.  The data strongly suggest that the
  local models, $Ll$ and $Lu$, belong to the LLI universality class
  whereas the nonlocal models, $Un$ and $Ln$, belong to new universality
  classes.}
\vspace*{0.5cm}
\begin{tabular}{lcccc}
Model   & $\tau$        & $\nu$         & $y$           & $\sigma$   \\
\tableline
$~Ll$   & $1.55\pm0.02$ & $2.24\pm0.03$ & $1.83\pm0.04$ & $1.42\pm0.03$ \\
$~Lu$   & $1.56\pm0.02$ & $2.26\pm0.03$ & $1.91\pm0.04$ & $1.37\pm0.03$ \\
$~Nl$   & $1.35\pm0.05$ & $1.55\pm0.05$ & $1.60\pm0.05$ & $0.95\pm0.05$ \\
$~Nu$   & $1.63\pm0.02$ & $2.75\pm0.05$ & $2.20\pm0.04$ & $1.50\pm0.04$ \\
\end{tabular}
\label{tab_exp}
\end{table}

\end{multicols}


\begin{thebibliography}{99}
\bibitem{other_functions}
Other functional forms can be used for $p(\delta_i)$.  Our
results show that as long as $p(\delta_i)$ converges rapidly
to the value one the same results are obtained.
\end{thebibliography}
\end{document}